\documentclass{revtex4}

\usepackage{epsfig}
\usepackage{amssymb}
\usepackage{graphicx}

\begin{document}
\title{ Quantum ballistic experiment on antihydrogen fall }
\author{A. Yu. Voronin$^{1,2}$, V.V. Nesvizhevsky$^3$, G. Dufour$^4$,  S. Reynaud$^4$}
\affiliation{ $^1$ P.N. Lebedev Physical Institute, 53 Leninsky
prospect, 117924 Moscow, Russia.
\\
$^2$ Russian Quantum Center, 100 A, Novaya street, Skolkovo, 143025, Moscow, Russia.
\\
$^3$ Institut Max von Laue - Paul Langevin (ILL), 71 avenue des Martyrs, F-38042, Grenoble, France.
\\
$^4$ Laboratoire Kastler Brossel, UPMC-Sorbonne Universités, CNRS, ENS-PSL Research University, Collège de France, Campus Jussieu, 75252, Paris, France.
}

\begin{abstract}

We study an interferometric approach to measure gravitational mass of antihydrogen ($\mathrm{\bar{H}}$). The method consists of preparing a coherent superposition of quantum state of $\mathrm{\bar{H}}$ localized near a material surface in the gravitational field of the Earth, and then observing the time distribution of annihilation events followed after the free fall of an initially prepared superposition of $\mathrm{\bar{H}}$ from a given height to the detector plate. We show that a corresponding time distribution is related to the momentum distribution in the initial state that allows its precise measurement. This approach is combined with a method of  production of a coherent superposition of gravitational states by inducing a resonant transition using oscillating gradient magnetic field. We estimate an accuracy of measuring the gravitational mass of the $\mathrm{\bar{H}}$ atom which could be deduced from such a measurement. 

\end{abstract}

\maketitle

\section{Introduction}
Antihydrogen atom ($\mathrm{\bar{H}}$), being the antimatter counterpart to Hydrogen atom, is of particular interest due to the simplicity of its internal structure, stability (in vacuum) and neutrality. These properties open possibilities for precision tests of CPT and the Equivalence Principle (EP). Tests of EP are especially interesting in case of quantum motion of antiatoms. Detailed studies of gravitational properties of antimatter are planned to be performed by most groups involved in experiments with $\mathrm{\bar{H}}$ \cite{Alpha,AlphaGrav, Yam, gabr10, cesa05, Aegis1}. 

Such experiments are extremely challenging. On one hand, this is due to the weakness of gravitational force compared to electromagnetic interactions, which means that careful elimination of any false effects related to electromagnetic forces is required. On the other hand, this is due to the very limited number of cold $\mathrm{\bar{H}}$ atoms available in the nearest future experimental set-ups. We study a typical case of a ballistic experiment in which an initially prepared $\mathrm{\bar{H}}$ state is released at a moment $t_0$ to fall down to the detector plate installed at a height downstream. The time distribution of annihilation events in the detector plate is related to the initial velocity and position distribution of antiatoms and is peaked around a mean value, corresponding to the classical time of fall. We show that shaping the vertical velocities in such an initial state, in order to reduce corresponding uncertainty, improves the accuracy of free-fall time measurement in spite of statistics reduction. In the ultimate limit, such a decrease of uncertainty in velocity and spatial distributions meets quantum limitations. 

Here we discuss a possibility of exploiting quantum properties of $\mathrm{\bar{H}}$ atoms in the gravitational field in order to measure the gravitational mass of $\mathrm{\bar{H}}$. Such quantum properties manifest themselves in the fact of existence of so-called gravitational states of $\mathrm{\bar{H}}$ \cite{GravStates}. These are long-living quantum states of $\mathrm{\bar{H}}$ in the gravitational field of the Earth above a material surface. In spite of naive expectation that any contact with material surface would result in prompt annihilation, such states have lifetimes of the order of fraction of a second due to the phenomenon of quantum (over-barrier) reflection of ultraslow antiatoms from steep antiatom-surface potential.

Shaping the vertical velocities can be performed by transmitting antiatoms through a slit between a bottom mirror and an upper absorber. We show that in the ultimate case of the slit size of a few dozens micrometers only a few quantum gravitational states can pass through. The time distribution of free fall events in this case maps the momentum distribution in those gravitational states. The interference between different gravitational states, defined by the energy difference between gravitational states, can be visualised via the time distribution of free-fall events and enable us to evaluate the gravitational mass of $\mathrm{\bar{H}}$ with high accuracy using potentially very precise interferometric methods.

\section{Galileo-type classical ballistic experiment}

Classical ballistic experiment is designed to measure the gravitational free fall acceleration $g$ of $\mathrm{\bar{H}}$. It consists of preparing an initial state (in case of GBAR experiment \cite{Yam} this state of $\mathrm{\bar{H}}^+$ is settled in the ion trap), releasing $\mathrm{\bar{H}}$ atom from the trap at a moment $t_0$ (in the GBAR experiment this moment corresponds to a laser photo-detachment pulse $\mathrm{\bar{H}}^+ + \hbar \omega\rightarrow \mathrm{\bar{H}}+ \mathrm{e^+}$), and detecting the moments of annihilation events in the detector plate installed at a height $H$ below the trap. The measured quantity is the time distribution of free-fall events. Such a distribution is related to the initial phase-space distribution of $\mathrm{\bar{H}}$. 

To find this relation we assume that the classical probability density to find $\mathrm{\bar{H}}$ in a vicinity of a phase-space point $(z,v)$ (here $v$ labels the vertical velocity and $z$ labels the vertical position) is given by function $P(z,v)$, while the relation between the classical time 
to reach the plane $z=0$ for 
 initial values of vertical position $z$ and vertical velocity $v$ is given by a well-known expression:
\begin{equation}\label{Tfall}
t=\sqrt{2z/g+v^2/g^2} +v/g,
\end{equation}
where $g$ is a free fall acceleration.

We assume that the probability density $P(z,v)$ is peaked around $z_0=H$ and $v_0=0$, and the corresponding standard deviations $\sigma_z \ll H$ and $\sigma_v\ll \sqrt{2gH}$, so that in the first order:
\begin{equation}\label {deltaT}
t-t_f\approx (z-H)/v_f+v/g.
\end{equation}
Here $t_f=\sqrt{2H/g}$ is a classical time of free fall from the height $H$ with zero initial velocity, $v_f=\sqrt{2gH}$ is a classical velocity, gained by a body during free fall from the height $H$.

Using the above expressions we arrive to the free-fall event distribution as a function of $\tau=t-t_f$:
\begin{equation}
N(\tau)=g \int P(z, \tau g-gz/v_f) dz.
\end{equation}

The most typical case corresponds to independent spatial and velocity distributions of Gaussian type:
\begin{equation}\label{ClassDistr}
P(z,v)=\frac{1}{\pi \sigma_z \sigma_v} \exp\left(-\frac{(z-H)^2}{\sigma_z^2}\right)\exp\left(-\frac{v^2}{\sigma_v^2}\right).
\end{equation}
 
In this case the free-fall events distribution has a form:
\begin{equation}
N(\tau)=\frac{1}{\sqrt{\pi} \sigma} \exp\left(-\tau^2/\sigma^2\right),
\end{equation}
where $\sigma=\sqrt{\sigma_z^2/(2gH)+\sigma_v^2/g^2}$.

The ratio, which defines the relative accuracy is given by:
\begin{equation}
\delta=\sigma/t_f=\sqrt{\sigma_z^2/(4H^2)+\sigma_v^2/(2gH)}.
\end{equation}

As one can see from the above expression, a trivial way  of improving accuracy is  to increase the time of free fall by
 increasing $H$. However this simple way is limited by dramatic increase in the cost of experimental installation with the increase of its size, as well as due to difficulty in eliminating false systematic effects during longer time of fall.

Usually, the main contribution to the time of free fall uncertainty $\sigma$ comes from the spread in initial velocity distribution $\sigma_v$. In particular, for the design of GBAR experiment, this value could be estimated as $\sigma_v\approx 0.5$ m/s (at least at the early stage of experiment). The corresponding relative accuracy in the evaluation of $t_f$ is:
\begin{equation}\label{epsT}
\epsilon_t=\frac{\delta}{\sqrt{N_{tot}}}\approx \frac{\sigma_v}{\sqrt{2gH N_{tot}}},
\end{equation}
 where $N_{tot}$ is the total number of free fall events.

Corresponding accuracy in the evaluation of free fall acceleration $g=2H/t_f^2$ is:
\begin{equation}\label{epsG}
\epsilon_g=2\epsilon_t=2\frac{\sigma_v}{\sqrt{2gH N_{tot}}}.
\end{equation}

Selecting antiatoms with small enough vertical velocities from a certain range could provide a method to decrease the uncertainty $\sigma_v$ and to improve the desired accuracy of evaluation of the value $g$.

Such selection could be realized by passing $\mathrm{\bar{H}}$ through a slit of size $h$ between two plates, the lower plate plays a role of mirror, while the upper plays a role of absorber \cite{Shaping}. This approach is equivalent to the one used in experiments with ultra cold neutrons (UCNs) in order to select certain states of vertical motion of neutrons in the gravitational field \cite{nesv00,nesv03,EPJC,NeutrWaveGuide}. We will discuss the physical principle of operating of such a shaping device in case of antiatoms in the next section. Here we mention only that within a classical description antiatoms, released from a trap at the height of bottom mirror with the vertical velocity $|v|>2gh$
 hit the absorber and are lost. Thus only antiatoms with a typical spread in the vertical velocity $\sigma'_v \approx \sqrt{2gh}$ would pass through the shaping device. We assume 
  that $\sigma'_v \ll \sigma_v$.

\begin{figure}
\centering
\includegraphics[width=0.80\textwidth]{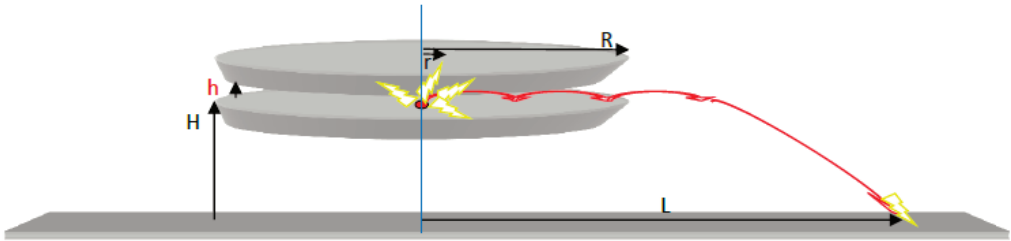}
\caption{\label{shaper} A sketch of experimental device for shaping velocity distribution of antiatoms. Antiatom trap is shown as a red circle in the centre of the shaping  disks. Bottom disk is a mirror, upper disk is an absorber. The disk radius is $R$, the trap radius $r\ll R$, the size of the slit between disks is $h$, the height of a bottom mirror above detector plate is $H$. Classical trajectories of antiatoms are shown by curved lines with arrows.}
\end{figure}

The sketch of a shaping device is shown in Fig.\ref{shaper}.
The total number of antiatoms which would be detected in the detector plate after passing the shaping device is reduced to $N'\approx N_{tot}\sigma'_v/\sigma_v$.
 Finally we get the following expression for the improved accuracy of measuring the value of $g$:
\begin{equation}
\epsilon'_g=2\frac{\sqrt{\sqrt{2gh} \sigma_v}}{\sqrt{2gHN_{tot}}}. 
\end{equation}

The improvement factor $\epsilon'_g/\epsilon_g$ is thus:
\begin{equation}
\frac{\epsilon'_g}{\epsilon_g}=\frac{\sqrt[4]{2gh}}{\sqrt{\sigma_v}}.
\end{equation}

One can see that shaping of vertical velocities results in  improvement of precision  in spite of the reduction of statistics. 

Let us mention that above semi-quantitative arguments are based on a classical description of $\mathrm{\bar{H}}$ dynamics. Improvement of velocity and position uncertainty of initial state by decreasing the slit size $h$ meets quantum restrictions for sufficiently small values of $h$. When the slit size $h$ becomes smaller than $h_q<50$ $\mu m$ account of quantum properties of antiatom motion in the gravitational field is essential. We will discuss in the following sections the method to utilize quantum properties for precision measurements of the gravitational mass of $\mathrm{\bar{H}}$.

\section{$\mathrm{\bar{H}}$ states near material surface and vertical velocity shaping.}

In this section we discuss in more details the physics behind $\mathrm{\bar{H}}$ interaction with the shaping device.

The bottom plate, being well-polished metallic surface, plays a role of mirror for $\mathrm{\bar{H}}$ atoms with vertical velocities of a few $cm/s$. It was shown in \cite{voro05l,voro05,GravStates,QRslabs} that the interaction of slow antiatoms with a material surface results in the reflection with a probability close to unity. This is explained by the effect of quantum (overbarrier) reflection of $\mathrm{\bar{H}}$ from the van-der Waals-Casimir-Polder potential between an (anti)atom and a material surface. 

\begin{figure}
  \centering
\includegraphics[width=100mm]{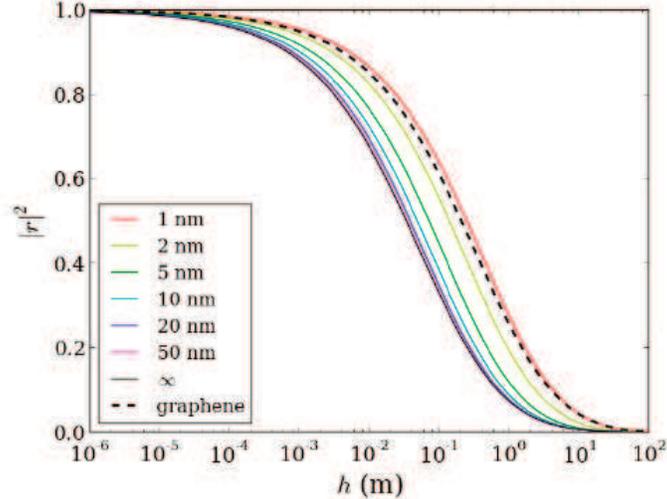}

\caption{The coefficient of reflection of the ground state $\mathrm{\bar{H}}$ atom from different surfaces, including silica
slabs of different width, as a function of the incident energy in units of the height of free fall.}\label{ThinSlabs}
\end{figure}

The $\mathrm{\bar{H}}$ reflection coefficient is given in Fig.\ref{ThinSlabs} as a function of incident energy.


In the limit of small incident velocities, the reflection coefficient is a linear function of $v$:
\begin{equation}
R=1-b_r v
\end{equation}
The factor $b_r$ is related to the scattering length $a_{CP}$ which characterizes the antiatom-surface interaction by 
\begin{equation}
b_r=4m |\mathop{\rm Im} a_{CP}|/\hbar.
\end{equation}

For the ideally reflecting surface it is equal to:
\[b_r=0.911 \mbox{ s/m}\]. 

The above law is valid, when $1-R\ll 1$. One can verify, it is well justified for velocities below $0.1 m/s$, which corresponds to the velocity gained during the free fall from the height $H=0.5$ mm. For such velocities the reflection coefficient is around 90\%. 

Thus a polished surface can play a role of a mirror for (anti)atoms with small enough vertical velocities. This means, in particular, that motion of $\mathrm{\bar{H}}$ in the gravitational field above such a reflecting surface is quantized (just like in case of UCNs \cite{Nature1, nesv00, nesv03}). Antiatoms are localized near a material surface in long-living gravitational states. The detailed study of such states can be found in \cite{GravStates}.

For convenience we summarize here the main results, concerning the gravitational states of $\mathrm{\bar{H}}$ above a material surface. In the following we distinguish between the gravitational mass, which we refer to as $M$ and the inertial mass, hereafter denoted by $m$. Confinement of $\mathrm{\bar{H}}$ atoms above a material surface in the gravitational field of the Earth is achieved due to the quantum reflection from the van der Waals- Casimir atom-surface potential. Such a potential has a large atom-wall distance asymptotic behavior $-C_4/z^4$, where $z$ is an atom-wall distance.

The characteristic length and energy scales of quantum states in the gravitational potential are given in the following expressions:

\begin{eqnarray}\label{scaleL}
l_g &=&\sqrt[3]{\frac{\hbar^{2}}{2mMg}}=5.871 \mbox{ }\mu m, \\
\varepsilon_g &=&\sqrt[3]{\frac{\hbar^2M^2g^2}{2m}}=2.211\cdot10^{-14} \mbox{ a.u.}.\label{scaleE}
\end{eqnarray}

The corresponding characteristic gravitational time scale is:
\begin{equation}\label{tauGrav}
\tau_g=\frac{\hbar}{\varepsilon_g}=0.001 \mbox{ s.}
\end{equation}

The typical spatial scale of the van der Waals/Casimir-Polder (further vdW/CP) potential is given by:
\begin{equation}
l_{CP}= \sqrt{2m C_4}/\hbar=0.003 \mbox{ } \mu m.
\end{equation}

The hierarchy of the vdW/CP interaction scale and the gravitational scale assumes that the energy levels and the corresponding wave-functions of gravitational states could be found using the ratio of the scattering length on vdW/CP potential and the gravitational length scale $a_{CP}/l_g\approx-i0.005$ as a perturbation parameter.

%
The energies and eigen-functions of gravitational states are given in the following equation:
\begin{eqnarray}\label{En}
E_n&=&\varepsilon_g (\lambda_n+a_{CP}/l_g), \\
\Phi_n(z)&=&C \mathop{\rm Ai}(z/l_g-\lambda_n-a_{CP}/l_g), \\
\mathop{\rm Ai}(-\lambda_n^0)&=0.
\end{eqnarray}
where $\mathop{\rm Ai}(x)$ is the Airy function \cite{abra72}, $\lambda_n$ are dimensionless roots of corresponding eigenvalue equations:
\begin{equation}
\frac{\mathop{\rm Ai}(-\lambda_n)}{\mathop{\rm Ai'}(-\lambda_n)}=-\frac{a_{CP}}{l_g}
\end{equation}


$\mathrm{\bar{H}}$ life time (the same for all lowest gravitational states) in case of the ideal conducting surface is:
\begin{equation}
\tau=\frac{m}{Mgb_r} \simeq 0.1 \mbox{s}. \label{time}
\end{equation}
This lifetime could be significantly larger for silica surface or thin slabs \cite{QRslabs}.

\begin{table}
 \centering
 \begin{tabular}{|c|l|l|l|l|}

 \hline
  $n$ & $\lambda_n^0$ &  $E_n^0$, peV & $E_n^0$, Hz & $z_n^0$, $\mu m$\\
  \hline
 1 & 2.338 &  1.407 & 340.321& 13.726  \\
 2 & 4.088 &  2.461 & 595.259 & 24.001  \\
  3 & 5.521 &  3.324 & 803.999 & 32.414    \\
  4 & 6.787 &  4.086 & 988.309& 39.846    \\
  5 & 7.944 &  4.782  & 1156.66 & 46.639        \\
  6 & 9.023 &  5.431  & 1313.63 &52.974      \\
  7 & 10.040&  6.044  & 1461.90 & 58.945   \\
  \hline
 \end{tabular}
 \caption{The eigenvalues, gravitational energies in peV and Hz, the classical turning points for a quantum bouncer with a mass of $\mathrm{\bar{H}}$ in the Earth's gravitational field.
 } \label{Table1}
 \end{table}


$\mathrm{\bar{H}}$ states between two horizontal plates, separated by a height $h$, differ from pure gravitational states due to the presence of the upper plate. The equation for the eigen-values, which accounts for the interaction with the gravitational field and with both surfaces has the following form:

\begin{equation}\label{EigenValues2}
\frac{\mathop{\rm Ai}(-\lambda_n+a_{CP}/l_g)}{\mathop{\rm Bi}(-\lambda_n+a_{CP}/l_g)}=\frac{\mathop{\rm Ai}(h/l_g-\lambda_n+a_{CP}/l_g)-a_{CP}/l_g \mathop{\rm Ai'}(h/l_g-\lambda_n+a_{CP}/l_g)}{\mathop{\rm Bi}(h/l_g-\lambda_n+a_{CP}/l_g)-a_{CP}/l_g \mathop{\rm Bi'}(h/l_g-\lambda_n+a_{CP}/l_g)}.
\end{equation} 
Here $\mathop{\rm Bi}(z)$ is the Airy function of second kind \cite{abra72}. Life-times of the corresponding states differ significantly depending whether the classical turning point meets the condition $\lambda_n l_g<h$. If this condition is true, the state is weakly pertubed by the upper plate and nearly coincides with a corresponding gravitational quantum state. In the opposite case, the state is strongly affected by second plate and the absorbtion takes place in both plates.
In the ultimate limit of higly excited states  $\lambda_n l_g\gg h$ the influence of gravity can be neglected and the state is close to the so called box-like state. Their energy and width is given by the following equation:
\begin{eqnarray}
E_n =\frac{\hbar^2\pi^2
n^2}{2m (h-2 a_{CP})^2}\\
 \Gamma_n\approx 4\left|\mathop{\rm Im} a_{CP}\right|\frac{\hbar^2\pi^2 n^2}{2m h^3}\label{ebox}.
\end{eqnarray}

The width of such states is significantly larger than the width of lower excited gravitational states. 

This effect explains state selective properties of a wave-guide, which consists of two parallel horizontal planes. States with quantum number $n$, such that $\lambda_n l_g<h$, mostly pass through the wave-guide, while states with higher $n$, such that $\lambda_n l_g>h$, are efficiently absorbed. In order to increase state selectivity, the upper plate surface is made rough. Non-grazing antiatom scattering on rough surface mixes vertical and much larger horizontal velocity components, which results in more effective absorption. The details can be found in \cite{NeutrWaveGuide}. 

\begin{figure}
 \centering
\includegraphics[width=140mm]{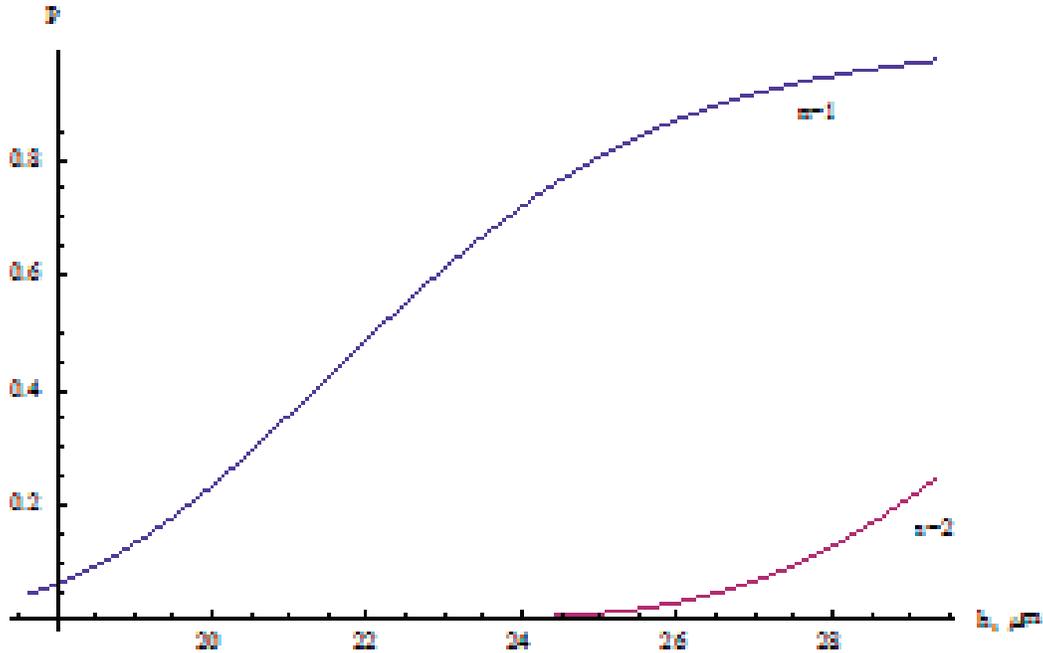}
\caption{Probability to find $\mathrm{\bar{H}}$ in a given state in the shaping device, as a function of the slit size $h$. The length of shaping device is 10 cm.}\label{StateSelect}
\end{figure}

We show the probability to find $\mathrm{\bar{H}}$ in the ground and first excited states in the shaping device as a function of the slit size $h$ in Fig. \ref{StateSelect}. One can see that positioning upper plate at the height of $14<h<24$ $\mu m$ results in selection of pure ground state.


%

\section {Free fall of a superposition of gravitational states.}

We would be interested below in studying the quantum regime of a ballistic experiment, when a shaping device selects one or a few lowest gravitational states. 

\begin{figure}
 \centering
\includegraphics[width=100mm]{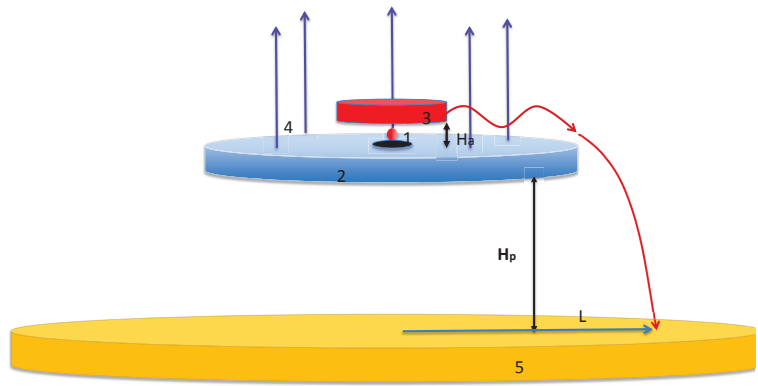}
\caption{A sketch of the principle scheme of an experiment on free fall of a superposition of $\mathrm{\bar{H}}$ gravitational states. 1 - a source of ultracold antihydrogen, 2 - a mirror, 3 - an absorber, 4 - a magnetic field, 5 - a detector plate.}\label{FigSketch4}
\end{figure}

A scheme of such an experiment is explaned below. Time zero is the moment of release of $\mathrm{\bar{H}}$ atoms from the trap. In GBAR experiment this moment is defined by a short photo-detachment laser pulse. After passing through the mirror-absorber shaping device, $\mathrm{\bar{H}}$ falls down from the edge of the mirror to the horizontal detection plate, positioned at a height $H_p=30$~cm
 below the mirror, as shown in Fig. \ref{FigSketch4}.
Now the detector plate is a position-sensitive detector, which allows one to measure the horizontal velocity of antiatoms via the relation $V_h=L_d/T$, where $L_d$ is a radial position of an annihilation event on the detection plate, $T$ is a time of flight starting from the moment of release. The value of horizontal velocity enables one to calculate the time spent by the atom on the mirror before falling down $t=L/V=LT/L_d$. With this correction one can find the time distribution of free-fall events. We are going to demonstrate that such a distribution is closely related to the momentum distribution $F_0(p)$ in an "initial"  wave-packet of $\mathrm{\bar{H}}$ atoms taken at the moment when such a wave-packet is at the edge of the mirror.
We suppose that the wavepacket is initially centred around the origin $z=0$ and calculate the flux through a detector plate in $z=-H_p$.
 The wave-packet free-fall evolution is determined using the free-fall propagator $G(p,p',t)$:
\begin{eqnarray}\label{FreeFallEv}
\Psi(z,t)=\frac{1}{\sqrt{2\pi \hbar}}\int_{-\infty}^{\infty} \int_{-\infty}^{\infty} G(p,p',t)F_0(p')\exp\left(ipz/\hbar\right)) dp dp',\\ \label{GFF}
G(p,p',t)=\exp\left(-\frac{it}{2m\hbar}(p^2+Mgpt+M^2g^2t^2/3)\right) \delta(p+Mgt-p').
\end{eqnarray}

The combination of both expressions gives the following:
\begin{equation}\label{GFF1}
\Psi(z,t)=\frac{1}{\sqrt{2\pi \hbar}}\int_{-\infty}^{\infty}  \exp\left(-\frac{it}{2m\hbar}(p^2+Mgpt+M^2g^2t^2/3)\right)F_0(p+Mgt)\exp\left(ipz/\hbar\right) dp , 
\end{equation}

For sufficiently large times of free fall $t\gg\tau_g$  the integral (\ref{FreeFallEv}) could be estimated using the stationary phase method, which gives:
\begin{equation}\label{FFstat0}
\Psi(z,t)\simeq \frac{1}{\sqrt{2\pi \hbar}}
\exp\left( -\frac{i g^2 M^2 t^3}{24 m \hbar }-\frac{i g M t z}{2 \hbar }+\frac{i m z^2}{2 t \hbar } \right)
 \int_{-\infty}^{\infty}\exp\left(-\frac{i t}{2\hbar m}(p-p_0)^2\right) F_0(p_0+Mgt) dp,
\end{equation}
where $p_0=mz/t-Mgt/2$
 is a stationary phase point.


For $t$ sufficiently large the integral in the right hand part of the above expressions converges for $p-p_0\lesssim \sqrt{2\hbar m/t}$. If this convergence momentum turns out to be much smaller than the  characteristic width $\delta k$ of the momentum distribution $F_0(k)$, so that  $ \sqrt{2\hbar m/t}\ll \delta k$, the function $F_0(k)$ can be taken out of integral.
 In this case we get:
\begin{equation}\label{FFstat}
\Psi(z,t)\simeq \sqrt{\frac{m}{i t}}\exp\left( -\frac{i g^2 M^2 t^3}{24 m \hbar }-\frac{i g M t z}{2 \hbar }+\frac{i m z^2}{2 t \hbar } \right) F_0 (p_0+Mgt),
\end{equation}

The above expression maps the initial momentum distribution $F_0(k)$ into the time- or position- distribution for large times $t$. In particular it means that a time distribution $P$ of free-fall events of a wave-packet, made of a coherent superposition of a few gravitational states (or only one gravitational state) is determined by a velocity distribution in the initial wave-packet, taken at the moment when this wave-packet started the free fall. This distribution is given by a flux through a detector plate:
\begin{eqnarray}\label{pt}
P&=&\frac{ \hbar}{2i m} \left( \Psi \frac{d\Psi^*}{dz}-\Psi^* \frac{d\Psi}{dz}\right)\Big|_{z=-H_p}\\
P&=&\left(\frac{M g}{2}+\frac{m H_p}{t^2} \right)\left|F_0\left(\frac{M g t}{2}-\frac{m H_p}{t}\right)\right|^2\simeq Mg|F_0\left(Mg(t-t_f)\right)|^2.
\end{eqnarray}
Here $H_p$ is a height of free fall, $t_f$ is a classical time of free fall from height $H_p$.

\begin{figure}
 \centering
\includegraphics[width=80mm]{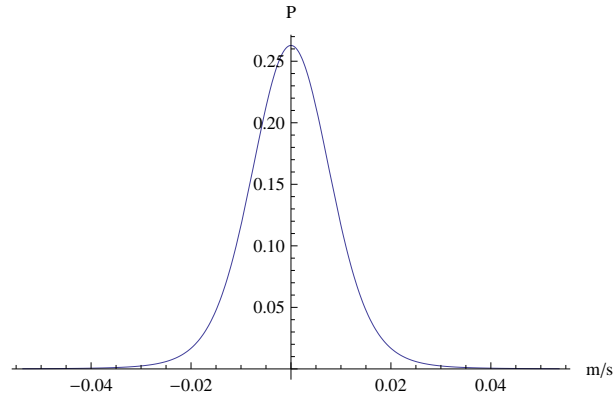}
\caption{The velocity distribution in a ground gravitational state.}\label{FigVelocity}
\end{figure}

\begin{figure}
 \centering
\includegraphics[width=80mm]{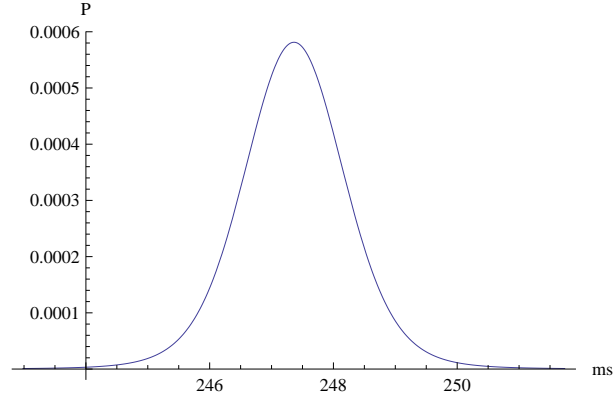}
\caption{The distribution of the time of free-fall events of a wave-packet, comprised of a ground gravitational state. The height of free fall is $H_p=30$ cm.}\label{FigTime}
\end{figure}

The distribution of initial velocities in a ground gravitational state has a characteristic width $\sigma_v=\sqrt{2 \epsilon_g/m}\approx 1.6$ cm/s. Such a distribution is shown in Fig.\ref{FigVelocity}, the corresponding arrival time distribution is shown in Fig.\ref{FigTime}. The  position of a maximum in the time-of-fall distribution is equal $t_f= 247.4$ ms, that corresponds to a free fall time $t_f=\sqrt{2mH_p/(Mg)}$ from the height of $30$ cm. The width of the distribution is $\delta=1.9$ ms, that gives an accuracy of measuring $M/m$ with $1000$ annihilation events of the order of $10^{-4}$.
The velocity distribution of a ground gravitational state  is more than an $30$ times less than initial velocity distribution of $\mathrm{\bar{H}}$ atoms, released from  GBAR trap. The corresponding increase of accuracy of $g$ measurement, with account of loss of statistics is one order of magnitude  compared to experiment without velocity shaping.

However, the use of quantum gravitational states opens much wider possibilities  by means of careful studying of time of arrival distribution in case a superposition of few gravitational states is used as initial wave-packet:
\begin{equation}
\Psi(z,t_0)=Ag_i(z)+B g_f(z).
\end{equation}
Here $t_0$ is the moment when free fall starts, $A,B$ are the amplitudes of gravitational states, $g_{i,f}(z)$ are the gravitational state wave-functions. 

\begin{figure}
 \centering
\includegraphics[width=80mm]{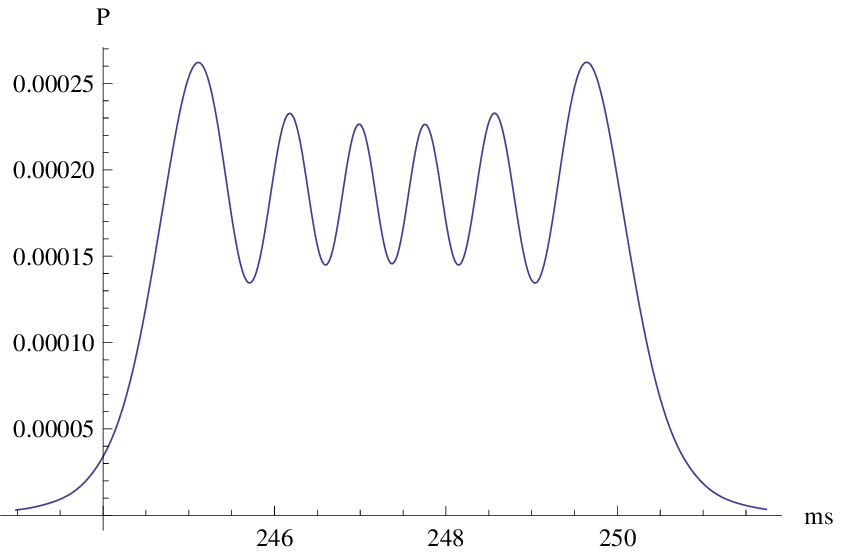}
\caption{The distribution of times of free-fall events for a wave-packet, comprised of a superposition of first and sixth  gravitational states. The height of free fall is $H_p=30$ cm. The amplitudes are $A=0$, $B=1$.}\label{FigFreefall6}
\end{figure}

\begin{figure}
 \centering
\includegraphics[width=80mm]{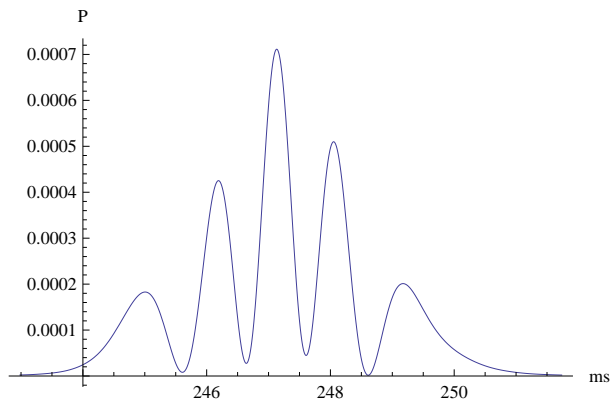}
\caption{The distribution of times of free-fall events for a wave-packet, comprised of a superposition of first and sixth  gravitational states. The height of free fall is $H_p=30$ cm. The amplitudes are $A=B=1/\sqrt{2}$. }\label{FigFreefall2}
\end{figure}

The distribution of arrival time events for a superposition of two gravitational states (ground state $g_1$ and sixth state $g_6$) is illustrated in Fig.\ref{FigFreefall6} and Fig.\ref{FigFreefall2} with different sets of amplitudes of gravitational states in both cases. As one can see the rich interference picture of initial velocity distribution is mapped in the arrival time distribution. It strongly depends on the relative amplitudes of gravitational states. 


In order to get useful information out of this interference picture one should control the relative amplitudes of gravitational states in their initial superposition. This can be done by means of inducing transition from initially prepared pure ground state into superposition of states by resonant gradient magnetic field. The details of inducing resonant transitions with inhomogeneous oscillating magnetic  are discussed in \cite{Voro14resgrav}. Here we will be interested in using analogous method for producing a superposition of gravitational states with controlled relative phase. 

An external magnetic field applied to the shaping device has the following form:
\begin{equation}\label{Magn1}
\vec{B}(z,x,t)=B_0 \vec{e}_z+ \beta \cos(\omega t) \left(z \vec{e}_z-x \vec{e}_x \right).
\end{equation}
Here $B_0$ is the amplitude of a constant, vertically aligned, component of magnetic field, $\beta$ is the value of magnetic field gradient, $z$ is a distance measured in the vertical direction, $x$ is a distance measured in the horizontal direction, parallel to the surface of a mirror.

An inhomogeneous magnetic field couples the spin and the spatial degrees of freedom. An $\mathrm{\bar{H}}$ wave-function is described in this case using a four-component column (in a nonrelativistic treatise) in the spin space, each component being a function of the c.m. coordinate $\vec{R}$, relative $\bar{p}-\bar{e}$ coordinate $\vec{\rho}$ and time $t$.

The interaction $\widehat{H}_m$  which is responsible for the field-magnetic moment interaction is:
\begin{equation}\label{Hm}
\widehat{H}_m=-2\vec{B}(z,x,t)\left( \mu_{\bar{e}}\hat{s}_{\bar{e}}\times \hat{I}_{\bar{p}}+ \mu_{\bar{p}}\hat{s}_{\bar{p}}\times \hat{I}_{\bar{e}}\right).
\end{equation}
Here $\mu_{\bar{e}}$ and $\mu_{\bar{p}}$ are magnetic moments of the positron and the antiproton respectively, $\hat{s}_{\bar{e}}$, $\hat{s}_{\bar{p}}$ is a spin operator, acting on spin variables of positron (antiproton), $\hat{I}_{\bar{e}}$, $\hat{I}_{\bar{p}}$ is a corresponding identity operator.

As far as the field $\vec{B}(z,x,t)$ changes in space and in time, this term couples the spin and the c.m. motion.

We will be interested in the weak field case, such that the Zeeman splitting is much smaller than the hyperfine level spacing $\mu_B B_0\ll \alpha_{HF}$. Here $\alpha_{HF}=2.157\cdot 10^-7$ a.u.. The hierarchy of interactions
\begin{equation}\label{hierar}
m_2e^2/\hbar^2 \gg \alpha_{HF}\gg \mu_{\bar{e}} |B_0 |\gg \varepsilon_n
\end{equation}
justifies the use of the adiabatic approximation; it is based on the fact that an internal state of an $\mathrm{\bar{H}}$ atom follows adiabatically the spatial and temporal variations of an external magnetic field. Neglecting non-adiabatic couplings, an equation system for the amplitude $C_n(t)$ of a gravitational state $g_n(z)$ has the form:
\begin{equation}\label{Adiab}
i \hbar \frac{d C_{n}(t)}{dt}= \sum_{k} C_{k}(t) V(t)_{n,k}\exp \left(-i\omega_{n k} t\right ).
\end{equation}
The transition frequency $\omega_{n k}=(\varepsilon_k-\varepsilon_n)/\hbar$ is defined by the gravitational energy level spacing. 

Within this formalism the role of the coupling potential $V(z,t)$ is played by the energy of an atom in a fixed hyperfine state thought of as a function of (slowly varying) distance $z$ and time $t$.
\begin{equation}
V(t)_{n,k}=\int_0^\infty g_n(z) g_{k}(z) E(t,z)dz.\\
\end{equation}
Here $g_n(z)$ is the gravitational state wave-function, which is given using the Airy function \cite{GravStates}.

The energy $E(z,t)$ is the eigenvalue of the internal and magnetic interactions, which depend on the c.m. coordinate $\vec{R}$ and time $t$ are treated as slow-changing parameters.
Corresponding expressions for the eigen-energies of a $1S$ manifold are:

\begin{eqnarray}\label{Ea}
E_{a,c}(z,t)&=&E_{1s}-\frac{\alpha_{HF}}{4}\mp\frac{1}{2}\sqrt{\alpha_{HF}^2+|(\mu_B-\mu_{\bar{p}})B(z,t)|^2},  \\ \label{Eb}
E_{b,d}(z,t)&=&E_{1s}+\frac{\alpha_{HF}}{4}\mp \frac{1}{2}|(\mu_B+\mu_{\bar{p}})B(z,t)|.
\end{eqnarray}

Subscripts $a,b,c,d$ are standard notations for hyperfine states of a $1S$ manifold in a magnetic field. The presence of a constant field $B_0$ produces the Zeeman splitting between states $b$ and $d$. As far as the energy of states $b,d$ depends on magnetic field linearly, while for states $a,c$ it depends quadratically, only transition between $b,d$ states takes place in case of a weak field. In the following we will consider only transitions between gravitational states in a $1S( b,d)$ manifold.

A qualitative behaviour of the transition probability is given in the Rabi formula, which can be deduced by means of neglecting the high frequency terms compared to the resonance couplings of only two states, initial $i$ and final $f$, in case the field frequency $\omega$ is close to the transition frequency $\omega_{if}=(E_f-E_i)/\hbar$:
\begin{eqnarray}\label{Rabbi}
\Psi(z,t)\sim \left[ \left(1+e^{i\Delta \omega t/2 }\left(\cos(\Omega t/2)-i\frac{\Delta\omega}{\Omega}\sin(\Omega t/2) \right)\right)g_i(z)-\frac{i V_{if}}{\hbar \Omega}e^{-i\Delta \omega t/2 }\sin(\Omega t/2)g_f(z)\right] e^{-\Gamma t/2}, \\
P_{if}=\frac{1}{2}\frac{(V_{if})^2}{(V_{if})^2+\hbar^2(\Delta \omega)^2}\sin^2\left(\frac{\sqrt{(V_{if})^2+\hbar^2(\Delta \omega)^2}}{2\hbar}t\right)\exp(-\Gamma t).
\end{eqnarray}
Here $\Delta \omega$ is resonance detuning, $\Omega=\sqrt{V_{if}^2/\hbar^2+\Delta \omega^2}$, $P_{if}$- transition probability from initial to final state during time $t$. The factor $1/2$ appears in front of the right-hand side of the above expression due to the fact that only two $(b,d)$ of four hyperfine states participate in the magnetically induced transitions.

In order to produce a superposition of ground and six-th gravitational state one needs to apply magnetic field with the resonance transition frequency equal to $\omega=972.46$ Hz. The value of the field gradient, optimized to obtain the maximum probability of $1\rightarrow 6$ transition during the time of flight $t_{fl}=\tau=0.1$ s, turned to be equal $\beta=27.2$ Gs/m, the corresponding guiding field value, which guarantees the adiabaticity of the magnetic moment motion, is $B_0=30$ Gs. 

Thus a magnetic field induced  transition opens possibility to prepare the state superposition with controllable amplitudes of gravitation states.

An experiment consists of releasing $\mathrm{\bar{H}}$ atoms at time $t_0$, selecting ground gravitational state while passing through the slit between mirror and absorber, positioned at the height $14<h<24$ $\mu$m above the mirror, preparing a superposition of ground and excited state by applying oscillating magnetic field to $\mathrm{\bar{H}}$ atoms moving above the surface of the mirror in the gravitational field, and finally detecting the time of arrival events on the detector plate, installed at the height $H_p=30$ cm below the mirror. The detector plate is also a position sensitive detector, which allows localizing the position of annihilation events and thus measuring the horizontal (classical) velocity of $\mathrm{\bar{H}}$ atoms and thus the time spent by an antiatom in the shaping device, as shown in Fig. \ref{FigSketch4}. 

The interference method of measuring the gravitational mass of $\mathrm{\bar{H}}$ is based on the fact, that the velocity distribution in a wave-packet, comprised of a coherent superposition of two gravitational states, includes several narrow peaks. These peaks are reproduced in the time-of-fall distribution. The positions of the two narrowest peaks, shown in Fig.\ref{FigFreefall2}, are equal $t_1=246.9$ ms and $t_2=247.8$ ms, with each width equal to $0.5$ ms. The corresponding relative uncertainty is $\varepsilon\approx 2\cdot 10^{-3}$.
The relative positions of these peaks provide an access to the characteristic momentum distribution of a gravitational state via Eq.(\ref{pt}). Namely, the knowledge of  maxima positions in the time distribution of free-fall events $t_m$ could be related to the position of corresponding maxima in the momentum distribution $p_m=\hbar k_m/l_g$, where $k_m$ mean dimensionless values:
 \begin{equation}
 \frac{\hbar k_m}{l_g}=Mg(t_m-t_0).
 \end{equation}
 This fact gives another access to experimental measurements of a characteristic energy scale $\varepsilon_g=Mgl_g$:
 \begin{equation}
 \varepsilon_g=\frac{\hbar k_m}{t_m-t_0},
 \end{equation}
 and thus to the gravitational mass of antihydrogen:
 \begin{equation}
M=\sqrt{\frac{2m\hbar k_{m}^3 }{g^2(t_m-t_0)^3}}.
\end{equation}
With $10^3$ annihilation events this approach provides a relative accuracy of the order of $10^{-4}$ for the value of $M$.

The method is especially powerful, as it provides the maximal detailed information. One can control the frequency of applied magnetic field, monitor the momentum distribution of a superposition of states mapped in the time of arrival distribution, and get these results for atoms with different horizontal velocities, and thus with different times spent in the shaping device.   


\section{Conclusion}
We propose a novel approach to study gravitational properties of antiatoms based on the interferometry of an initially prepared superposition of quantum states of $\mathrm{\bar{H}}$ in the gravitational field of the Earth near a material surface. The interference pattern is mapped in the time-of-arrival annihilation signal, obtained when such a superposition is dropped from the edge of the mirror down to the detector plate. The study of such a time distribution allows measuring the mean time of fall, as well as details of the momentum distribution of initial states superposition. We showed that forming initial coherent state is possible by means of passing the flux of $\mathrm{\bar{H}}$ atoms through a shaping device, which consists of a material horizontal mirror and an absorber installed above it at the height of $14<h<24$ $\mu$m. In spite of unavoidable loss of statistics this procedure improves accuracy of measurements due to significant decrease of the width of velocity distribution of initial state. The coherent superposition of two gravitational states is formed by inducing resonant transition between gravitational states by oscillating gradient magnetic field. We showed that measuring the positions of narrow interference maxima (minima) in the time distribution of annihilation signal makes possible the measurement of the gravitational mass $M$ of the $\mathrm{\bar{H}}$ atom with relative accuracy of order of $10^{-4}$ with $1000$ annihilation events.


\end{document}